# Gamification and AI: Enhancing User Engagement through Intelligent Systems

Carlos J. Costa
Advance/ISEG (Lisbon School of Economics & Management), Universidade de Lisboa,
Lisbon, Portugal
cjcosta@iseg.ulisboa.pt

Joao Tiago Aparicio
LNEC, INESC-ID, Instituto Superior Técnico,
Universidade de Lisboa,
Lisboa, Portugal
Joao.aparicio@tecnico.ulisboa.pt

Manuela Aparicio
NOVA Information Management School (NOVA IMS), Universidade Nova de Lisboa, Portugal
manuela.aparicio@novaims.unl.pt

Sofia Aparicio
Instituto Superior Técnico,
Universidade de Lisboa,
Lisboa, Portugal
sofia.aparicio@tecnico.ulisboa.pt

**Abstract**

Gamification applies game mechanics to non-game environments to motivate and engage users. Artificial Intelligence (AI) offers powerful tools for personalizing and optimizing gamification, adapting to users' needs, preferences, and performance levels. By integrating AI with gamification, systems can dynamically adjust game mechanics, deliver personalized feedback, and predict user behavior, significantly enhancing the effectiveness of gamification efforts. This paper examines the intersection of gamification and AI, exploring AI's methods to optimize gamified experiences and proposing mathematical models for adaptive and predictive gamification.

## 1. Introduction

Gamification motivates users to complete tasks through points, badges, challenges, and other rewards (Deterding et al., 2011). While gamification has proven effective in various fields such as education, marketing, and workplace productivity, integrating AI enables a more nuanced approach, with systems that respond to individual user behavior and dynamically adjust incentives and feedback (Hamari et al., 2014). The combination of gamification and AI promises a new level of engagement with personalized user experiences, adaptive progression, and optimized rewards. In recent years, gamification has emerged as a powerful strategy in various domains, including education, marketing, and healthcare, by applying game design elements to non-game contexts (Zichermann & Cunningham, 2011). This approach aims to enhance user engagement, motivation, and retention by leveraging intrinsic and extrinsic motivators (Ryan & Deci, 2000). As educational platforms seek innovative ways to boost learner involvement and success, the integration of gamification principles becomes increasingly relevant (Kapp, 2012).

However, despite its growing popularity, gamification strategies are often challenging, particularly when addressing users' diverse needs and preferences (Koivisto & Hamari, 2019). This paper explores the intersection of gamification and artificial intelligence (AI). We investigate how AI can enhance gamification by personalizing experiences and adapting to users' behavior in real time (Bontchev, Terzieva & Paunova-Hubenova, 2021). This leads us to the central research question: How can AI-driven gamification be effectively modeled and implemented in a learning platform to optimize user engagement and retention? This study aims to propose a framework for AI-driven gamification that integrates mathematical modeling techniques to create a more personalized and adaptive learning experience. To achieve this, we employ a multi-faceted approach that includes theoretical exploration of gamification principles (Deterding et al., 2011), application of AI techniques (Ocumpaugh et al., 2024), and development of a mathematical model to analyze and predict user behavior.

The structure of this paper is as follows: Gamification and Key Components, Artificial Intelligence in Gamification, Mathematical Modeling of AI-Driven Gamification, and Case Study: AI-Driven Gamification in a Learning Platform. So, the following section overviews gamification, discussing its foundational elements and how they contribute to user engagement. In the following section, we examine the role of AI in enhancing gamification, focusing on personalization, adaptation, and predictive analytics (Kizilcec et al., 2013). The following section presents a mathematical framework that models gamification dynamics in the context of AI, emphasizing the relationship between user engagement, rewards, and retention. The case study section describes a practical application; we illustrate the proposed framework through a case study of a learning platform that utilizes AI to optimize gamification strategies. By examining these interconnected components, this paper aims to provide insights into the effective integration of AI and gamification, offering valuable implications for research and practice in educational technology.

## 2. Gamification and Key Components

Gamification typically includes rewards, progression, challenges, and feedback. These elements motivate users to engage and provide feedback on their performance and behavior. Key components include Reward Systems, Adaptive Progression, Challenges and Feedback, and Retention and Motivation.

Reward Systems incentivize users with points, badges, and achievements. This component is crucial for maintaining user engagement and motivation throughout the gamified experience (Deterding et al., 2011).

Adaptive progression involves levels that increase in difficulty as users progress, ensuring that the challenge remains appropriate to the user's skill level and maintains a state of flow (Csikszentmihalyi, 1990).

Challenges and feedback personalize challenges and provide real-time feedback that adjusts to user performance. This component is essential for creating a sense of accomplishment and guiding users toward improvement (Kapp, 2012).

Retention and motivation allow sustaining user engagement by balancing rewards with meaningful challenges. This balance is critical for long-term user retention and continued motivation (Hamari et al., 2014).

The integration of these components creates a gamification usage ecology, as Costa et al. (2017) described. This ecology encompasses the interplay between user motivations, game mechanics, and the overall context of the gamified system.

## 3. Artificial Intelligence in Gamification

Artificial Intelligence (AI) brings several benefits to gamification, including data-driven personalization, predictive modeling, and real-time adaptability. Integrating AI in gamification represents a significant advancement in creating more engaging and compelling user experiences (Costa et al., 2023). Key AI methodologies in gamification include personalization with machine learning, predictive analytics, and adaptive game mechanics.

### 3.1 Personalization with Machine Learning

Machine learning algorithms analyze user data, such as activity patterns, preferences, and performance, to create a personalized experience. By clustering similar user profiles or using reinforcement learning, AI can adjust the difficulty, type of challenges, and rewards, maintaining an optimal level of engagement (Costa et al., 2024). This personalization is crucial in maintaining user interest and motivation over time.

Example: A learning platform using AI may adjust quiz difficulty based on user accuracy, motivation, and engagement, ensuring users remain within their "flow zone"—challenging enough to be engaging without causing frustration. This adaptive approach aligns with the concept of "flow" in gamification, where the challenge level matches the user's skill level (Aparicio et al., 2019).

### 3.2 Predictive Analytics

Predictive models use historical data to anticipate user behavior. For instance, AI can predict which users are at risk of disengagement and trigger specific interventions, such as additional rewards or reminders. These models use regression analysis, Markov chains, and neural networks (Costa & Aparicio, 2023). Predicting and preemptively addressing potential disengagement is a powerful tool in maintaining long-term user participation.

Example: A fitness app might predict a drop in user activity based on past behavior patterns. It could then use gamified incentives, such as time-limited badges, to encourage continued engagement. This proactive approach to user retention is particularly effective in health and wellness applications (Aparicio et al., 2021).

### 3.3 Adaptive Game Mechanics

AI enables systems to adjust mechanics in real-time. Adaptive mechanics involve using reinforcement learning to reward actions that encourage desired behaviors, optimizing engagement. For instance, if a user loses motivation, AI may reduce task difficulty or increase rewards to keep them engaged (Costa et al., 2017). This dynamic adjustment of game elements ensures that the gamification system remains challenging and rewarding for users at all skill levels.

Example: In an educational game, AI may adjust hints and feedback frequency based on user performance, making tasks easier for struggling users and more challenging for advanced users. This adaptive approach enhances learning outcomes and maintains a high level of engagement across diverse user groups (Piteira et al., 2018).

## 4. Mathematical Modeling of AI-Driven Gamification.

We can apply several mathematical models to structure and optimize AI-driven gamification components, such as adaptive rewards, progression, and user engagement. These models provide a framework for implementing personalized and engaging gamification experiences (Costa et al., 2024). Below, we introduce some key formulas and models.

## 4.1 Dynamic Reward Systems

AI models can optimize reward distribution by calculating ideal reward frequency and value, adapting based on user engagement and difficulty (Hamari et al., 2014; Landers et al., 2017).

Reward Frequency: An exponential function represents a system where reward intervals increase as users engage more, promoting sustained interaction::

$$R(t) = R_0 \cdot e^{\alpha t}$$

Where:

$R(t)$ represents rewards at time t,

$R_0$ is the initial reward frequency,

$\alpha$ adjusts reward scaling based on engagement data.

**Diminishing Rewards:** To prevent over-rewarding, diminishing returns can be applied to reduce reward value over time:

$$V(n) = \frac{V_0}{1 + \beta n}$$

Where:

$V(n)$ represents reward value after n interactions,

$V_0$ is the initial reward value,

$\beta$ adjusts the rate of diminishing returns.

## 4.2 Adaptive Progression and Difficulty Scaling

Progression systems benefit from models that align task difficulty with user skill level, providing an experience that's neither easy nor overly challenging (Kapp, 2012; Csikszentmihalyi, 1990)..

**Logistic Progression Model**: This function can adjust task difficulty based on user performance, creating adaptive progression:

$$D(x) = \frac{D_{max}}{1 + e^{-\gamma(x - x_0)}}$$

Where:

$D(x)$ Is difficulty at the level $(x)$

$D_{max}$ is the maximum difficulty level

$\gamma$ controls the rate of progression

$x_0$ is the baseline skill level

**Flow Theory Model**: To keep users in an optimal flow state, a linear relationship between task difficulty C(x) and skill level s(x) can be maintained:

$$C(x) = s(x) + k$$

Where:

    C(x) is challenge level at user skill x,

    s(x) represents user skill level,

    k is a constant maintaining balance.

### 4.3 Predictive Engagement and Retention Models

AI can predict and optimize engagement and retention rates by modeling how likely a user is to remain engaged over time (Costa & Aparicio, 2023; Deterding et al., 2011)..

**Retention Probability:** A logistic regression model estimates retention probability P(t) as a function of time t, engagement factors E, and reward factors R :

$$P(t) = \frac{1}{1 + e^{-(aE+bR-c)}}$$

Where:

    *a* and *b* are coefficients representing engagement and reward impact,

    c is a threshold parameter.

**Engagement Decay:** Engagement often decays over time. An exponential decay function models this decline, with AI adjusting engagement factors in response:

$$E(t) = E_0 \cdot e^{-\lambda t}$$

Where:

    E(t) represents engagement at time t,

    $E_0$ is initial engagement,

    $\lambda$ is the decay constant.

These mathematical models provide a foundation for implementing AI-driven gamification systems that adapt to individual user needs and behaviors (Zichermann & Cunningham, 2011; Aparicio et al., 2019). By leveraging these models, developers can create more engaging and effective gamified experiences that maintain user interest over time (Seaborn & Fels, 2015; Sailer et al., 2017).

### 5. Case Study: AI-Driven Gamification in a Learning Platform

This case study examines a learning platform that leverages artificial intelligence (AI) and gamification to enhance user retention and engagement. The platform employs AI models to personalize the learning experience, providing tailored challenges, predictive retention analytics, and adaptive rewards based on individual user behavior.

In order to improve user engagement, the system continuously monitors several key metrics, including user accuracy, engagement levels, and time spent on tasks. This data informs a dynamic difficulty adjustment mechanism that alters task challenges according to user performance, ensuring that challenges remain appropriately aligned with each learner's skill level.

A central platform component is the logistic regression model, which predicts the likelihood of user retention. By analyzing engagement and reward values, the model identifies users at risk of disengagement, allowing the platform to implement timely interventions such as motivational messages or personalized rewards. Additionally, a dynamic reward system adjusts the frequency and magnitude of rewards based on user performance and engagement, effectively mitigating the risk of diminishing returns from repetitive tasks.

The architecture of the AI-Driven Gamification Platform is built on three key principles: engagement, skill acquisition, and reward mechanisms. It utilizes a logistic regression model to determine user retention probabilities based on critical metrics, facilitating the identification of users needing additional support.

The Core Components of the Platform are dynamic difficulty adjustment, reward mechanism, engagement tracking, and retention prediction.

Dynamic Difficulty Adjustment feature employs a logistic progression model, adjusting task difficulty based on user success rates to maintain an optimal challenge level. In the reward mechanism, an exponential decay model is applied to reward values, reflecting the principle of diminishing returns in order to sustain user motivation over time. Engagement Tracking allows continuous tracking of user interactions with gamified elements and allows for timely adjustments to the learning experience. Retention prediction uses a logistic regression model to aid in predicting retention probability, pinpointing users who may be at risk of dropping out.

The platform is implemented as a Python application utilizing libraries such as scikit-learn (Pedregosa et al., 2011) for machine learning, `numpy` for numerical operations, and matplotlib (Hunter, 2007) for data visualization. Below are the program's key components: data preparation, logistic regression model, visualization, and simulated learning session.

The program generates synthetic data representing user engagement, task performance, and reward metrics. This data is split into training and testing sets to evaluate the logistic regression model.

```python
import numpy as np
from sklearn.model_selection import train_test_split

# Generate synthetic data
np.random.seed(0)
num_samples = 1000
engagement = np.random.rand(num_samples)
rewards = np.random.rand(num_samples) * 10
retention = (engagement * 0.5 + rewards * 0.5 > 5).astype(int)  # Simple retention criterion

# Split data into training and testing sets
X_train, X_test, y_train, y_test = train_test_split(
```

```
        np.column_stack((engagement, rewards)), retention, test_size=0.2, random_state=42
)
```

The logistic regression model is trained on the prepared data to predict retention probabilities.

```
from sklearn.linear_model import LogisticRegression
from sklearn.metrics import accuracy_score, confusion_matrix

# Train logistic regression model
model = LogisticRegression()
model.fit(X_train, y_train)

# Predict on test set
y_pred = model.predict(X_test)
accuracy = accuracy_score(y_test, y_pred)
cm = confusion_matrix(y_test, y_pred)
```

The results, including the retention probabilities and the model's performance metrics, are visualized using `matplotlib`.

```
import matplotlib.pyplot as plt

# Plot confusion matrix
plt.figure(figsize=(8, 6))
plt.imshow(cm, interpolation='nearest', cmap=plt.cm.Blues)
plt.title('Confusion Matrix')
plt.colorbar()
tick_marks = np.arange(2)
plt.xticks(tick_marks, ['Not Retained', 'Retained'])
plt.yticks(tick_marks, ['Not Retained', 'Retained'])
plt.ylabel('True label')
plt.xlabel('Predicted label')
plt.show()
```

The program simulates a learning session by dynamically adjusting task difficulty and rewards based on user performance and engagement levels.

```
def simulate_learning_session(num_tasks):
    for task in range(num_tasks):
        engagement = np.random.rand()
```

```
        rewards = np.random.rand() * 10

        difficulty = calculate_difficulty(engagement, rewards)

        success = np.random.rand() < (1 - difficulty)  # Simulate success based on difficulty

        print(f'Task {task + 1}: Engagement: {engagement:.2f}, Reward: {rewards:.2f}, Difficulty: {difficulty:.2f}, Success: {success}')

def calculate_difficulty(engagement, rewards):
    return 1 / (1 + np.exp(-(engagement + rewards - 1)))

# Run simulation
simulate_learning_session(10)
```

The following charts show the engagement, skill level, reward, difficulty level, and retention probability over time.

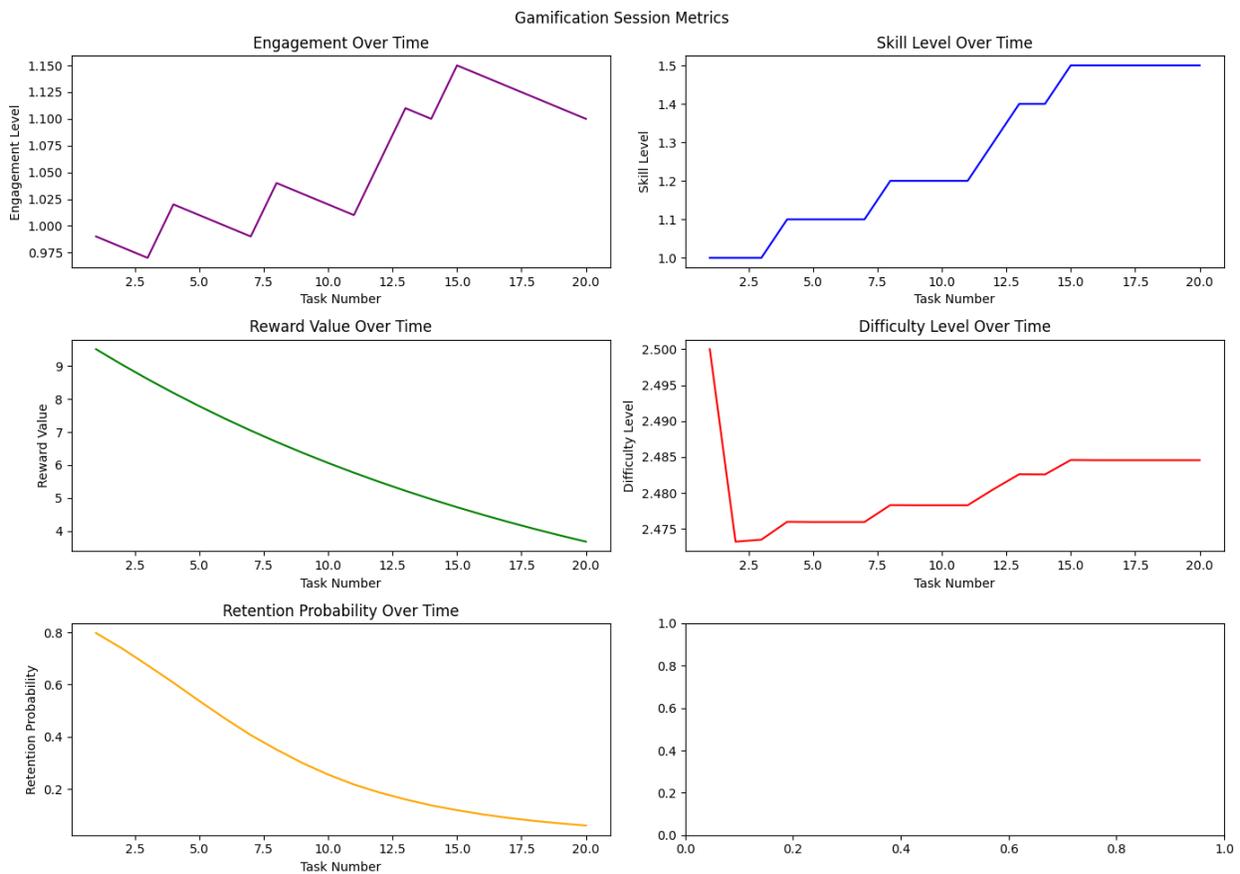

The insights gained from the simulation and machine learning analysis help refine the learning experience offered by the platform: Engagement Dynamics, Skill Growth, Reward System Efficiency, and Retention Strategy.

Users show varying engagement levels, which can be monitored and analyzed to identify when they are most likely to disengage. Tracking skill levels over time illustrates the effectiveness of the gamification strategy in fostering skill acquisition. Analyzing the impact of the reward system reveals how well it motivates users and encourages task completion. The logistic regression model provides actionable insights into user retention, allowing for targeted interventions to keep users engaged.

This improved section incorporates a detailed description of the Python program, outlining its structure, the libraries used, and the key functionalities implemented. It connects mathematical modeling with practical application, providing a comprehensive view of how the platform operates and how it leverages data to enhance user engagement and retention.

**6. Conclusion and Future Directions**

Combining gamification with AI provides new avenues for creating engaging, responsive, and highly personalized systems. Mathematical models allow for structured approaches to reward optimization, engagement prediction, and adaptive progression. Future research should develop more sophisticated AI models that refine real-time adaptation and response to user behavior, possibly integrating deep learning and real-time data analytics to enhance user experience.

Integrating AI with gamification provides transformative potential for dynamic, adaptive, and engaging experiences across sectors. Mathematical models provide a foundational framework to optimize gamification, ensuring it remains effective, personalized, and sustainable.

By integrating AI-driven methodologies and mathematical modeling, the learning platform demonstrates the potential of gamification to enhance user experience. The ongoing data collection and analysis allow for continuous system improvement, ensuring it remains responsive to user needs and preferences. This case study exemplifies how gamification, backed by data-driven insights, can transform educational experiences and foster lasting engagement.